# Reply to 'Comment on "Relativistic shape invariant potentials"'

A. D. Alhaidari

Physics Department, King Fahd University of Petroleum & Minerals, Box 5047,

Dhahran 31261, Saudi Arabia

E-mail: **haidari@mailaps.org**

The points raised in the Comment are addressed and except for one error, which will be corrected, the conclusion is that all of our findings are accurate.

The Hamiltonian that resulted in the radial equation (1) of our Paper [1] is not the minimum coupling Hamiltonian $H$ shown on the second page of the Paper but the one obtained from it by replacing the two off-diagonal terms $\alpha \vec{\sigma} \cdot \vec{A}$ with $\pm i\alpha \vec{\sigma} \cdot \vec{A}$, respectively. Consequently, our interpretation of $(V, \hat{r}W)$ as the electromagnetic potential and the statement that "$W(r)$ is a gauge field" are not correct. Likewise, calling equation (3) in the Paper, or any other derived from it, as the "gauge fixing condition" is not accurate. This has to be replaced everywhere by the term "constraint". Nevertheless, aside from an error which is corrected below, all developments based on, and findings subsequent to equation (1) still stand independent of that interpretation.

One of the main contributions in our Paper is the choice of constraint, which resulted in Schrödinger-like equation for the upper spinor component. This makes the solution of the relativistic problem easily obtainable by correspondence with well-known exactly solvable nonrelativistic problems. As such, we find enough justification for making that particular choice of constraint as given by equation (4). Furthermore, the angle parameter $\rho$ in the unitary transformation (2) is proportional to $\alpha$ as shown explicitly on page 9830 when taking the nonrelativistic limit of the Dirac-Rosen-Mörse I potential (i.e., $\tan 2\rho \approx \alpha\tau$). Therefore, this transformation does reduce to the identity in the nonrelativistic limit ($\alpha \to 0$).

The error, which was pointed out in the Comment [2] of assigning the inadmissible value $\kappa = 0$ in the cases where $V(r) = 0$, was hastily made to eliminate the centrifugal barrier $\kappa(\kappa+1)/r^2$ and the term $2\kappa W/r$ simultaneously from equation (11) so that we end up with the "super-potential" $W^2 - W'$. This mistake, which will now be corrected, affects only the Dirac-Rosen-Mörse II, Dirac-Scarf, and Dirac-Pöschl-Teller problems. Eliminating these two terms can be achieved properly by replacing the potential function $W(r)$ given in the Paper for each of the three problems by $W(r) - \kappa/r$, where $\kappa$ is now arbitrary. That is, in equations (10) and (11) of the Paper and in the Table we substitute the following potential function for the corresponding problem:

Dirac-Rosen-Mörse II: $W(r) = F\coth(\lambda r) - G\csch(\lambda r) - \kappa/r$

Dirac-Scarf: $\qquad\qquad W(r) = F\tanh(\lambda r) + G\sech(\lambda r) - \kappa/r$

Dirac-Pöschl-Teller: $\quad W(r) = F\tanh(\lambda r) - G\coth(\lambda r) - \kappa/r$

where $F$, $G$, and $\lambda$ are the potential parameters defined in the paper. This gives the same differential equations for the spinor components and reproduces the same solutions (energy spectrum and wave functions) as those given in the Paper for each of the three problems.